# The Physics of Electric Field Effect Thermoelectric Devices

**V. Sandomirsky, A. V. Butenko, R. Levin[1] and Y. Schlesinger**
Department of Physics, Bar-Ilan University, Ramat-Gan 52900, Israel
[1]The College of Judea & Samaria, Ariel 44837, Israel

**ABSTRACT**

We describe here a novel approach to the subject of thermoelectric devices. The current best thermoelectrics are based on heavily doped semiconductors or semimetal alloys. We show that utilization of electric field effect or ferroelectric field effect, not only provides a new route to this problem, bypassing the drawbacks of conventional doping, but also offers significantly improved thermoelectric characteristics. We present here model calculation of the thermoelectric figure of merit in thin films of Bi and PbTe, and also discuss several realistic device designs.

**INTRODUCTION**

Application of an electric field (Electric Field Effect - EFE) to a capacitor structure of the type Gate-Dielectric-Semimetal (or non-degenerate semiconductor), or Gate-ferroelectric-Semimetal injects, into the sample, charge carriers which are distributed among the electron and hole bands. Depending on the field polarity, the number of carriers of the corresponding sign increases, while the number of carriers of the other sign diminishes. This is connected with the change of the electron ($\eta_e$) and the hole ($\eta_h$) Fermi levels with an applied electric field, i. e. $\eta_e=\eta_e(E)$ and $\eta_h=\eta_h(E)$. Increasing the electric field can result in zero concentration of the corresponding charge carrier type. A further increase of the electric field results in a further increase of the introduced carriers and hence of the conductivity. It follows from the above argument that all the thermoelectrical effects, such as Peltier, Seebeck, Nernst etc are strongly dependent on the magnitude of the electric field. Thus, the EFE offers a novel, controllable and more effective route to thermoelectricity. In the following we will term this sort of thermoelectrics by EFE-TE.

In following we analyze in detail the electric field dependence of the Seebeck effect [1]. One can show that the Seebeck effect is expressed by the formula

$$S = S_p - S_n, \qquad (1)$$

where

$$S_p = \frac{k_B}{e}\frac{\langle \sigma_p s_p \rangle}{\langle \sigma_p + \sigma_n \rangle}; \quad S_n = \frac{k_B}{e}\frac{\langle \sigma_n s_n \rangle}{\langle \sigma_p + \sigma_n \rangle}. \qquad (1')$$

Here $\sigma_p$ and $\sigma_n$ are the hole and the electron conductivities, $s_p$ and $s_n$ are the hole and the electrons related Seebeck coefficients, $<>$ denotes averaging over the sample thickness. The later is needed because of the inhomogeneous electric field in sample (the larger the dielectic constant $\varepsilon$ the larger the characteristic screening length $L_s$, determining the rate of the exponential decrease of the field in the sample).

In absence of electric field ($E=0$), $S_n$ and $S_p$ are of comparable magnitude, therefore $S$ is smaller than each of the above quantities. For a certain value of the mobilities, $S_p$ could equal $S_n$ and therefore $S=0$. It should be noted (Eq. 1') that $S$ depends on the electric field via the field dependence of $\sigma$ (due to a change of carrier concentration with $E$) and the field dependence of $s_{n,p}$ (via the dependence of the Fermi energies on $E$) according to

$$s_k = \left[ \frac{(\frac{5}{2}+\lambda)F_{\frac{3}{2}+\lambda}}{(\frac{3}{2}+\lambda)F_{\frac{1}{2}+\lambda}} - \eta_k \right], \qquad (2)$$

where

$$\eta_k = \frac{(\varepsilon_F)_k}{k_B T}; \qquad F_i = \int_0^\infty \frac{x^i\, dx}{1+\exp(x-\eta)}.$$

Here $\lambda$ is the exponent in the energy dependence of the scattering relaxation time $\tau$ ($\tau \propto \varepsilon^\lambda$), $\eta_k$ is the reduced Fermi level (index $k$ indicating the hole or the electrons related quantities), and $F_i$ is the Fermi integral.

The unique characteristic of the method we propose here, lays in its ability to dynamically optimize and vary the thermoelectric behavior of the system (by varying the applied electric field magnitude and polarity) under a variety of conditions.

**TEST-CASE CALCULATIONS**

**a.** *Bi films.*

The curves of Fig. 1 were calculated assuming a value of film thickness $L=100$ , and temperature $T=300$ K. It should be noted that the detailed shape of the curve in Fig. 1 (in particular, via the energy dependence of the mobilities) depends also on the particular scattering mechanism (e.g. scattering by defects, ionized impurities, acoustic and/or optical phonons etc).

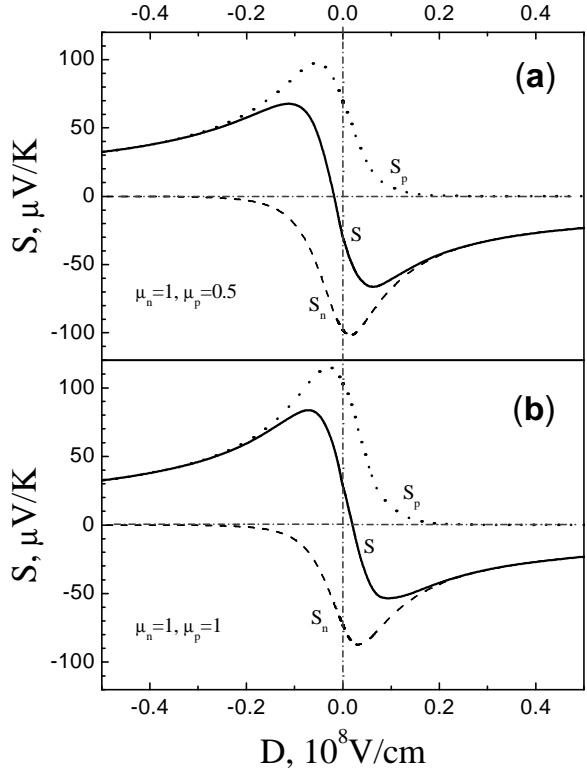

**Figure 1.**

The total, $S$, and the partial ($S_p$ due to holes an $S_n$ due to electrons) Seebeck coefficients as a function of the electric displacement $D$.
(a) - mobilities $\mu_n=1$, $\mu_p=0.05$ m$^2$/Vsec
(b) - mobilities $\mu_n=1$, $\mu_p=1$ m$^2$/Vsec

The curves have been calculated assuming only electron-acoustic phonon scattering.

With increasing positive polarity (for definiteness) the hole concentration, and therefore also $S_p$ goes to zero. At the same time $|S_n|$ goes through a maximum and then falls with increasing $E$ (because of the increase of $\eta$ with $E$). A similar behavior (with the roles of electrons and holes interchanged) is obtained for negative $E$. This is demonstrated in Fig. 1 by the two extrema of $S(D)$ ($D$ is the electrical displacement) at $D_{m,pos}$ and $D_{m,neg}$.

The quality of the thermoelectric behavior is usually characterized by the dimensionless, sample averaged, figure of merit [2]

$$M = \frac{\langle \sigma S \rangle^2}{\langle \sigma \rangle \kappa} T, \qquad (3)$$

where $\sigma$ is the electrical conductivity.

The electric field dependence of $M$ is shown in Fig. 2. The two maxima of $M$ are related to the two extrema of $S$ (see Fig. 1). They appear at slightly shifted values of $E$, due to the monotonical field dependence of $\sigma$. It follows thus, that (together with the field variation of the thermoelectric quantities described above) also a strong field dependence of $M$ should be observed. This leads to the possibility that thin Bi films (which are not good thermoelectrics) can actually attain a high $M$ under electric field conditions.

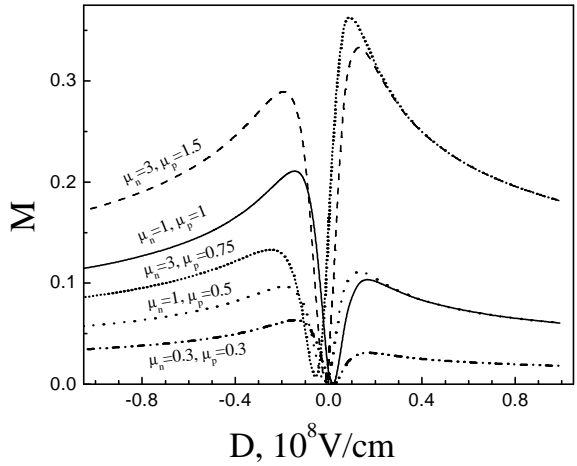

**Figure 2.**

The sample averaged figure of merit, for different combinations of the electron and hole mobilities, as a function of the electric displacement *D*.

The curves have been calculated assuming only electron-acoustic phonon scattering.

**b.** *PbTe film.*

The currently typical thermoelectrics are the, narrow gap, doped semiconductors $A_{IV}B_{VI}$ such as $Bi_2Te_3$, PbTe etc. Fig. 3 shows the calculation results of the maximum value of *M(D)*, for PbTe, as a function of film thickness *L*. The range of film-thickness *L*<700 was chosen so that $L \leq L_s$ (the screening length), assuring a reasonable homogeneity of the electric field over the sample thickness.

The present "worst case" calculations are based upon an assumed value of mobility $\mu = 0.05$ m$^2$/Vsec (a typical value determined by the current thin film preparation methodology) and assuming electron-acoustic phonon scattering mechanism only. Even so, despite the low mobility and the unfavorable energy dependence of $\tau$ ($\tau_{e-ph} \propto \varepsilon^{-0.5}$), the calculated figure of merit for PbTe under the electric field effect, reaches a considerably high value *M*≥2 for a thin (*L*<120 ) film, see Fig. 3. In analogy with the technological advances

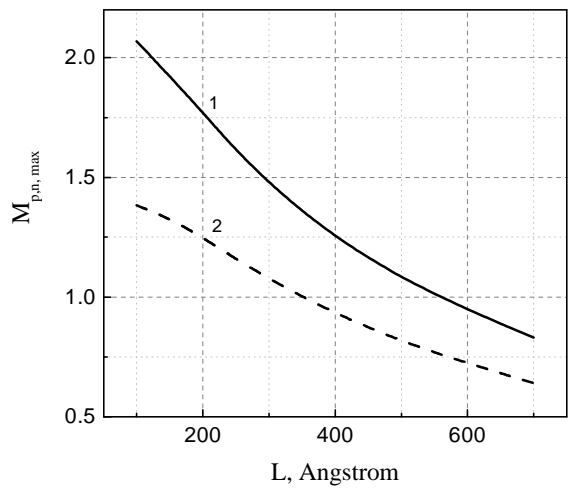

**Figure 3.**

The maximum sample averaged figure of merit *M(E)* for thermoelectric layer of
(1) - negative polarity (p-type)
(2) - positive polarity (n-type)
as a function of film thickness *L*.

in Bi film preparation methods [3], we expect a similar improvement in the PbTe system, resulting in a higher mobility (actually, a value of $\mu = 0.2$ m$^2$/Vsec, at room temperature, has been reported just now [4]). Moreover, it is noteworthy that, in general, for a given value of mobility, $M(E)$ will be significantly higher when other scattering mechanisms (besides electron-acoustic phonon scattering) are involved.

**EFE-TE DEVICE STRUCTURAL CONFIGURATIONS**

The analysis and calculations, outlined in the previous sections, have been carried out assuming the simplest model structure (a parallel plate capacitor, comprising of a metallic gate, a dielectric layer and a semimetal or semiconductor film). However, it is clear that more complex structures can be developed for higher efficiency and/or for specific purposes. In the following we describe shortly several of the configurations currently studied, both experimentally as well as theoretically.

**a.** *p-n thermoelement.* This two-arm structure, shown in Figure 4, is obtained by replacing the metallic gate by a second layer of thermoelectric material. This device exploits the electric field effect mechanism, causing the negative polarity thermoelectric layer to become an n-type semimetal (or semiconductor) and, at the same time, the positive polarity layer to become p-type. This structure thus serves as a controllable thermoelectric analogue of the standard p-n element. The device can be optimized choosing different thermoelectrics for each arm.

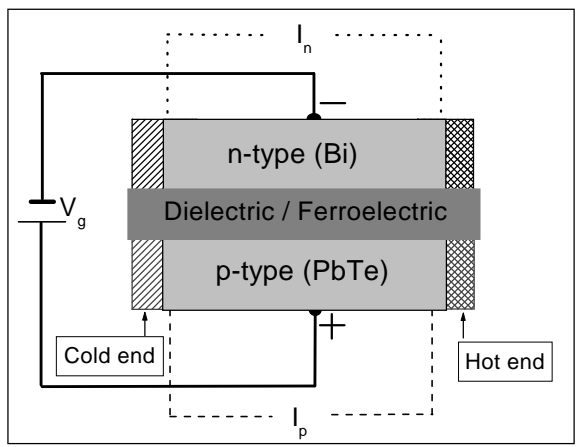

**Figure 4.**

The two-arm p-n thermoelement.

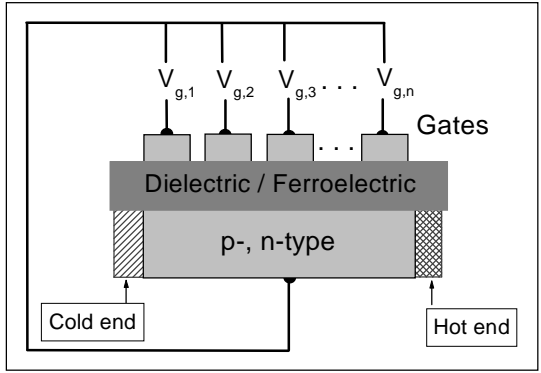

**Figure 5.**

The sectioned gate EFE-gradient thermoelement.

**b.** *EFE gradient thermoelectric device*. In this device, shown in Figure 5, a sectioned gate configuration or a resistive gate, create the conditions analogous to the gradient doping in conventional thermoelectrics. It should be emphasized that in the EFE-TE case the gradient can be dynamically varied, thus offering an active device.

**c.** *A split-gate Peltier cooler/heater device*. Here, as shown in Figure 6, opposite polarity voltage is applied to the two sections of the split gate thus creating a p-n junction. In such way, the direction of current determines whether the p-n junction absorbs or emits heat.

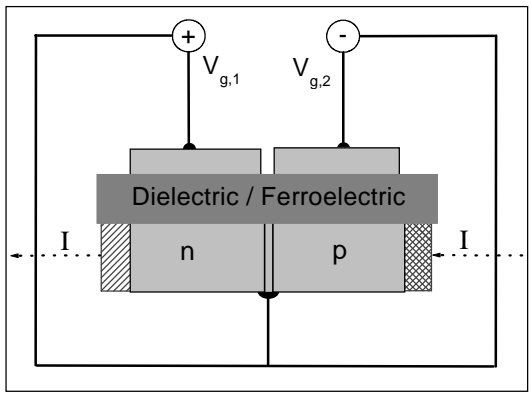

**Figure 6.**

The split-gate Peltier cooler/heater.
Cooling or heating functions are determined by the direction of the current.